\documentclass[prb,twocolumn,showpacs,amsmath,amssymb,amsfonts
]{revtex4}
\usepackage{graphicx}
\usepackage{graphics}
\usepackage{dcolumn}
\usepackage{bm}
\usepackage{amssymb}
\usepackage{amsmath}
\usepackage{amsfonts}
\usepackage{epsfig}
\newcommand {\el} {\\ \nonumber}
\newcommand {\bra} [1] {\langle #1 |}
\newcommand {\ket} [1] {| #1 \rangle}
\newcommand {\bkt} [1] {\langle #1 \rangle}
\newcommand {\dbkt} [2] {\langle #1 | #2 \rangle}
\newcommand {\tbkt} [3] {\langle #1 | #2 | #3 \rangle}
\newcommand {\pd} [2] {\frac{\partial #1}{\partial #2}}
\newcommand {\td} [2] {\frac{d #1}{d #2}}

 \newcommand {\beq}{\begin{equation}}
\newcommand {\eeq}{\end{equation}}
\newcommand {\bea}{\begin{eqnarray}}
\newcommand {\eea}{\end{eqnarray}}
\begin{document}
\title{Coherent wave-packet evolution in coupled bands}
\author{Dimitrie Culcer}
\author{Yugui Yao}
\author{Qian Niu}
\affiliation{Department of Physics, The University of Texas at
Austin, Austin TX 78712-1081} \affiliation{International Center
for Quantum Structures, Chinese Academy of Sciences, Beijing
100080, China}
\date{\today}
\begin{abstract}
We develop a formalism for treating {\it coherent} wave-packet
dynamics of charge and spin carriers in degenerate and nearly
degenerate bands. We consider the two-band case carefully in view
of spintronics applications, where transitions between spin-split
bands often occur even for relatively weak electromagnetic fields.
We demonstrate that much of the semiclassical formalism developed
for the single-band case can be generalized to multiple bands, and
examine the nontrivial non-Abelian corrections arising from the
additional degree of freedom. Along with the center of mass motion
in crystal momentum and real space, one must also include a
pseudo-spin to characterize the dynamics between the bands. We
derive the wave packet energy up to the first order gradient
correction and obtain the equations of motion for the real- and
k-space center of the wave-packet, as well as for the
pseudo-spin. These equations include the non-Abelian Berry
curvature terms and a non-Abelian correction to the group
velocity. As an example, we apply our formalism to describe
coherent wave-packet evolution under the action of an electric
field, demonstrating that it leads to electrical separation of up
and down spins. A sizable separation will be observed, with a
large degree of tunability, making this mechanism a practical
method of generating a spin polarization. We then turn our
attention to a magnetic field, where we recover Larmor precession,
which cannot be obtained from a single-band point of view. In this
case, the gradient energy correction can be regarded as due to a
magnetic moment from the self-rotation of the wave-packet, and we
calculate its value for the light holes in the spherical four-band
Luttinger model.
\end{abstract}
 \maketitle

\section{Introduction}
It often happens, in transport phenomena, that one has to consider
carrier dynamics in bands which are coupled together. This
coupling arises either through strong interband scattering or as a
result of the bands being degenerate, or both. The nearly
degenerate case is particularly relevant in transport theory as
transitions often occur between bands even at relatively weak
electromagnetic fields. Such situations include two-dimensional
systems described by the Rashba Hamiltonian\cite{Rashba} with
strong scattering, the doubly degenerate heavy and light hole
bands in the Luttinger model\cite{ChowKoch}, which is frequently
used to model the valence bands of bulk zincblende semiconductors,
and the conduction bands of wurtzite structures\cite{Cardona}. The
case of nearly degenerate bands has not, to date, received the
attention it deserves\cite{QiZhang}, despite the important role
played by such bands in semiconductor spintronics
systems\cite{Johnson, MolReview}, whether in dealing with spin
currents\cite{Spincurrent}, spin generation\cite{Spingeneration}
and relaxation \cite{Spinrelaxation}, or spin injection across a
semiconductor interface \cite{Spininjection}.

Spintronics systems lend themselves to a semiclassical treatment,
as the external electromagnetic fields vary on scales that are
considerably larger than atomic size. The semiclassical formalism
has had much success in describing carrier dynamics and transport
phenomena in condensed matter physics. In the non-degenerate case,
the carrier dynamics can be obtained semiclassically then combined
with the Boltzmann equation to produce accurate descriptions of
the transport properties of many materials. This approximation is
used in the descriptions of cyclotron orbits, conduction in
solids, the Hall effect and magnetoresistance\cite{AM}. An
essential application of the semiclassical model, which is
specifically relevant to our discussion, is in treating external
fields that are not represented by bounded operators, so that a
perturbative expansion will not converge\cite{Callaway}. The most
common example is provided by uniform electric and magnetic
fields, where the potential is linear in position.

We therefore develop, in this paper, a semiclassical description
of transport in degenerate and nearly degenerate bands. One of our
main purposes is to extend the semiclassical approach, as
developed by Sundaram and Niu\cite{Ganesh}, to the case of coupled
Bloch bands, in order to take into account the spin degree of
freedom. We illustrate the underlying physics by treating two
bands, without loss of generality. Two-band models are frequently
an adequate description of the conduction bands of many
semiconductors\cite{Twoband}. In experiments on spin transport in
semiconductors the carriers have traditionally been electrons
\cite{Eltransport}, as the strong spin-orbit coupling in the
valence band causes holes to lose spin information much
faster\cite{MolReview}. However, in recent years research has also
focused on spin currents in the valence bands of semiconductors
\cite{Spincurrent}, with a degeneracy which is usually greater
than two, and the formalism we outline is straightforwardly
extended to multiple bands.

To formulate a description of coherent transport in coupled bands
we may no longer work with each band individually but must instead
treat the coupled-band manifold as a whole. The condition for our
theory to be valid, which in the one-band case states that there
must be no transitions out of that band\cite{AM}, translates into
the requirement that there be no transitions out of the manifold
under consideration. We will consider a wave packet made up of two
bands, which is a suitable description of {\it coherent}
transport, when the density matrix has off-diagonal terms and the
relative phase of the two wave functions plays a crucial role.
This approach allows us to retain the notion of the real-space
center of the wave-packet, ${\bf r}_{\rm c}$, which remains well
defined. Moreover, in extending the formalism to two bands we are
able, in the presence of a magnetic field, to recover Larmor
precession, which is not possible from a one band picture. The
additional degree of freedom of the two-band system can be taken
into account by defining a wave function with the Bloch
periodicity in such a way as to incorporate both bands, which
allows us to derive the dynamics from a single-band point of view.
The coefficients of the bands can then be grouped into a vector
which we shall call the pseudo-spin, the structure and dynamics of
which makes clear the gauge structure of the problem. An
interesting fact which will emerge from our analysis is that the
effect of the external perturbations can be incorporated entirely
into the Berry curvatures \cite{Ganesh}, which in turn are
generated by a set of connections in real and reciprocal space as
well as in time. The Berry curvatures acquire additional terms
needed to ensure gauge covariance, and in the framework we present
they take the form of field strength tensors associated with the
connections.

The organization of this paper is as follows. In Section II we
develop the semiclassical formalism for coherent transport in the
presence of electromagnetic fields, deriving the Lagrangian, based
on a time-dependent variational principle, and the equations of
motion. In Section III we use our formalism to show how coherent
wave-packet evolution under the action of an electric field leads
to the separation of up and down spins. This idea is similar in
principle to the spin transistor proposed by Datta and Das
\cite{Das}. We demonstrate that a large degree of tunability can
be achieved by varying the gate field and number density. Finally,
in section IV we examine the case of a magnetic field. We show
that the gradient correction to the energy can be interpreted as
an intrinsic magnetic moment of the wave-packet \cite{Artem, Ganesh, Chang},
and we calculate this angular momentum correction for the light
holes in the spherical four-band model of the Luttinger
Hamiltonian.

\section{Formalism}
The semiclassical model describes the dynamics of wave packets.
The wave packet we consider is well localized in reciprocal space,
and it is assumed it sees only a small part of the lattice at any
one time. It is chosen in such a way that its spread in wave
vector is much smaller than the size of the Brillouin zone, so
that its motion at any moment is dependent only on the local
properties of the band structure. In order for this to happen, the
uncertainty principle dictates that the spread in real space must
be greater than the size of the lattice constant.

We consider systems whose Hamiltonians are functions of slowly
varying parameters, such as the potentials of weak external
electromagnetic fields, which vary on larger length scales than
that of the wave packet, and are treated classically. The periodic
potential of the ions on the other hand, changing over dimensions
small compared to the wave packet spread, must be treated quantum
mechanically \cite{AM}. 

Given these conditions, we define the {\it local} Hamiltonian ${\rm \hat H_c({\bf r}_c, t)}$ as the Hamiltonian with the slowly varying potentials evaluated at the center of the wave packet, which we denote by ${\rm{\bf r}_c}$, and time t. The Hamiltonian may be expanded \cite{Ganesh} about ${\rm {\bf r}_c}$ and if the external fields vary on spatial scales much larger than that of the wave packet we may truncate the expansion at the gradient term, which we define by ${\rm \Delta \hat H}$: \beq\label{DeltaH} \Delta \hat H = \frac{1}{2} [({\bf \hat r}-{\bf
r}_c) \cdot \frac{\partial \hat H_c}{\partial{\bf r}_c} +
c.c.].\eeq The gradient term gives rise to a
correction to the energy, which will play an important role in our
discussion below. 

The energy spectrum of the local Hamiltonian ${\rm \hat H_c}$ consists, as usual, of a series of bands, of which several are close together in energy and are separated from the others by larger gaps. It is the subset spanned by these bands that constitutes the focus of our attention. We regard the fields in this problem as small enough that Zener tunneling to the remote bands is negligible, but they may still be strong enough to induce transitions within the subset. For an energy spectrum with such a structure we may further decompose the local Hamiltonian into a degenerate part, ${\rm \hat H_d}$, which, when restricted to the subset of bands closely spaced in energy is proportional to the identity matrix, and a non-degenerate part, ${\rm \hat H_n}$, which is assumed small and treated perturbatively. The local Hamiltonian of (\ref{DeltaH}) is then:
\begin{equation}
\hat H_c = \hat H_d + \hat H_n.
\end{equation}
The gradient correction to ${\rm \hat H_c}$ can also be expressed in terms of the degenerate and non-degenerate contributions:
\begin{equation}
\Delta\hat H_c = \Delta\hat H_d + \Delta\hat H_n.
\end{equation}
Since ${\rm \hat H_n}$ is treated as a perturbation the gradient correction to it, ${\rm \Delta\hat H_n}$ will be second order in smallness. We will therefore neglect this correction henceforth. 

When the external fields are smoothly varying the states move within the subset of bands which are close in energy and which henceforth, for simplicity and without loss of generality, we take to be two-dimensional. 
The subset is spanned by two basis functions, which are
eigenstates of ${\rm \hat H_d}$, the degenerate part of the local Hamiltonian, evaluated at
${\rm {\bf r}_c}$, which has the periodicity of the unperturbed crystal:
\beq \hat H_d \ket{\Psi_i({\bf r}_c, {\bf q}, t)} = \epsilon\ket{\Psi_i({\bf r}_c, {\bf q}, t)}. \eeq For
a given ${\rm {\bf r}_c}$, therefore, these eigenstates have the Bloch
form, with the functions ${\rm \ket{u_i}}$ representing the lattice
periodic parts of the wave functions:
\begin{eqnarray}
|\Psi_1({\bf r}_c, {\bf q}, t)\rangle=e^{i{\bf q}\cdot{\bf \hat r}}|u_1({\bf r}_c, {\bf q},  t)\rangle \\
|\Psi_2({\bf r}_c, {\bf q}, t)\rangle=e^{i{\bf q}\cdot{\bf \hat
r}}|u_2({\bf r}_c, {\bf q}, t)\rangle
\end{eqnarray}
The wave functions ${\rm |u_i({\bf r}_c, {\bf q},  t)\rangle}$ are
spinors with the full periodicity of the lattice. Despite the fact
that the two bands are spin split, it cannot be assumed that their
local spin quantization axes are antiparallel, as the interactions
with neighboring bands may affect the direction of quantization.
Therefore, in principle, a finite overlap exists between the
spinors corresponding to the two bands and it is not revealing to
make a further decomposition of the eigenfunctions into an orbital
and a spin part. Additionally, the Hamiltonian contains terms
describing the spin-orbit interaction, which may depend on wave
vector and position.

Employing the crystal momentum representation, the wave packet is
expanded in the basis of Bloch eigenstates: \beq
\ket {w} =\int d^3q\{ a({\bf q}, t)[\eta_1({\bf q}, t)\ket{\Psi_1} +
\eta_2({\bf q}, t)\ket{\Psi_2}]\}. \eeq 
As the wave packet depends only
on the local properties of the band structure, the basis functions
$\ket{\Psi_1}$, $\ket{\Psi_2}$ are functions of the position
of the wave packet center, {\bf r}$_{\rm c}$, wave vector and time,
although implicit in the ket notation is dependence on position.
The function ${\rm a({\bf q}, t) = |a({\bf q},t)|e^{-i\Gamma({\bf
q},t)/2}}$, which incorporates the overall phase term, is a narrow
distribution function describing the extent of the wave packet in
reciprocal space and is sharply peaked at the center of the wave
packet, denoted by ${\rm {\bf q}_c}$, as discussed by Sundaram and
Niu \cite{Ganesh}. The functions $\eta_1$ and $\eta_2$ describe the
composition of the wave packet in terms of the two bands. 
The wave packet satisfies the normalization conditions:
\begin{equation}
\label{norm} \int d^3q |a|^2 = 1, |\eta_1|^2 + |\eta_2^2| =1.
\end{equation}
The wave packet can be rewritten by grouping together the
coefficients in an overall wave function ${\rm \ket{u}}$, which retains
the Bloch periodicity:
\begin{equation}
|w\rangle= \int d^3q|a|e^{-i\Gamma/2}e^{i{\bf q}\cdot{\bf \hat
r}}|u\rangle.
\end{equation}
Note that ${\rm \ket{u}}$ is not an eigenstate of the local Hamiltonian ${\rm \hat H_c = \hat H_d + \hat H_n}$, but
an expansion in eigenstates of ${\rm \hat H_d}$, a crucial difference from the
one-band situation. In addition, the wave vector and time dependence of ${\rm \ket{u}}$
come both from the time dependence of the Bloch states and that
of the coefficients.

We require the real space center of the wave packet to be given
by:
\begin{equation}
{\bf r}_c= \langle w| {\bf \hat r} |w\rangle = \frac{\partial
\Gamma_c}{\partial {\bf q}_c} + {\bf R}_c
\end{equation}
The subscript c signifies that the quantity is evaluated at the
center of the wave packet in reciprocal space, that is ${\bf q} =
{\bf q}_{\rm c}$. The vector ${\bf R}$, representing a connection in
reciprocal space, is defined as follows: \beq {\bf R} = \langle
u|i\frac{\partial}{\partial{\bf q}}|u\rangle .\eeq

The energy of the wave packet is given by the expectation value
\begin{equation}\begin{split}
\langle w| \hat H |w\rangle = \langle w| \hat H_d |w\rangle  +\langle w| \hat H_n |w\rangle  +
\langle w| \Delta \hat H_d |w\rangle \\ \equiv \epsilon + \Delta_n + \Delta_d \equiv \varepsilon.
\end{split}\end{equation} Both ${\rm \Delta_n}$ and ${\rm \Delta_d}$ are expressible entirely in terms of the Bloch wave function ${\rm \ket{u}}$. ${\rm \Delta_n}$ is given by 
\begin{equation}\begin{split}
\Delta_n = \langle u| \tilde H_n |u\rangle = \eta_i^* \Delta^n_{ij}\eta_j, \Delta^n_{ij}\eta_j = \langle u_i| \tilde H_n |u_j\rangle,
\end{split}\end{equation}
while ${\rm \Delta_d}$ is
\begin{equation}\begin{split}
\Delta_d = \frac{i}{2}(\tbkt{u}{\frac{\partial \tilde H_d}{\partial{\bf
r}_c}\cdot}{\pd{u}{{\bf q}}} - c.c.) - \frac{\partial \hat \epsilon}{\partial{\bf r}_c} \cdot {\bf R}.
\end{split}\end{equation}
In the above ${\rm \tilde H_n = e^{-i{\bf q}\cdot\hat{\bf r}} \hat H_ne^{i{\bf q}\cdot\hat{\bf r}}}$ while ${\rm\tilde H_d = e^{-i{\bf q}\cdot\hat{\bf r}} \hat H_de^{i{\bf q}\cdot\hat{\bf r}}}$. The energy correction ${\rm\Delta_d}$ is identical to the expression obtained by Sundaram and Niu \cite{Ganesh}. It takes on an additional significance when a magnetic field is present, as will be seen in the last section.

The Lagrangian ${\cal L}$ is obtained semiclassically by means of
a variational principle:
\begin{equation}
{\cal L} = \langle w|(i\hbar \frac {d} {dt} - \hat H)|w \rangle
\end{equation}
Its use is justified by the fact that the Euler-Lagrange equation
of motion for ${\rm |w\rangle}$ derived from it is the time-dependent
Schrodinger equation. Following the method used by Sundaram and
Niu, the following expression is found for the
Lagrangian: \begin{equation}\begin{split} {\cal L} = \langle
u|i\hbar\td{u}{t}\rangle + \hbar{\bf \dot r}_c\cdot{\bf q}_c - \varepsilon
= \\ = i\hbar \eta_i^*\td{\eta_i}{t} + \hbar \eta_i^* \dbkt{u_i}{i\td{u_j}{t}} \eta_j + \hbar{\bf \dot r}_c\cdot{\bf q}_c - \\ - \epsilon - \eta_i^*(\Delta^n_{ij} + \Delta^d_{ij})\eta_j.
\end{split}\end{equation} 
In the above ${\rm \td{}{t}}$ represents the total time derivative, including both the explicit time dependence and the implicit, which is due to dependence on ${\bf q}_{\rm c}$ and ${\bf r}_{\rm c}$. The Lagrangian depends only on the values of $\eta_{\rm i}$ and ${\rm \td{\eta_i}{t}}$ along the trajectory ${\rm {\bf q} = {\bf q}_c (t)}$. Since ${\bf q}_{\rm c}$ is a function of time only, we may regard $\eta_{\rm i}$ in the Lagrangian as an independent variable, ${\rm \eta_i(t)}$. The equations of motion derived from the Lagrangian are:
\begin{equation}\begin{split}
\hbar\dot{\bf q}_c = - \pd{\epsilon}{{\bf r}_c} +
({\bf \Omega_{rr}}\dot{\bf r}_c + {\bf \Omega_{rq}}\dot{\bf q}_c)
- {\bf \Omega}_{t{\bf r}}\\
 \hbar\dot{\bf r}_c = \pd{\epsilon}{{\bf
q}_c} - ({\bf \Omega_{qr}}\dot{\bf r}_c + {\bf
\Omega_{qq}}\dot{\bf q}_c) + {\bf \Omega}_{t{\bf q}}\\
i\hbar\td{\eta_i}{t} = ({\cal H}_{ij} - \hbar\dbkt{u_i}{i\td{u_j}{t}})\eta_j. 
\end{split}\end{equation} 
Here ${\rm {\cal H}_{ij} = \tbkt{u_i}{\hat H_c}{u_j}}$. The curvature tensor
${\rm \Omega_{{\bf rr}}^{\alpha\beta}}$ is defined by: \beq \Omega_{{\bf
rr}}^{\alpha\beta} =
i(\dbkt{\pd{u}{r_\alpha}}{\pd{u}{r_\beta}}-\dbkt{\pd{u}{r_\beta}}{\pd{u}{r_\alpha}})\eeq
and the vector ${\rm {\bf \Omega}_{t{\bf q}}}$ by:\beq \Omega_{t{\bf
q}}^\alpha =
i(\dbkt{\pd{u}{t}}{\pd{u}{q_\alpha}}-\dbkt{\pd{u}{q_\alpha}}{\pd{u}{t}})\eeq
The others can be deduced analogously. These quantities have
exactly the same form as the curvatures defined in the paper by
Sundaram and Niu \cite{Ganesh}.

We specialize in the case of an external electromagnetic field.
The effect of such an external field is discussed thoroughly by
Sundaram and Niu \cite{Ganesh}. The wave vector {\bf q} must be
replaced by ${\rm {\bf k} = {\bf q} + \frac{e}{\hbar}{\bf A} ({\bf r},
t)}$, which is the gauge invariant crystal momentum (for electrons
with charge -e), and therefore the Hamiltonian will have the
form ${\rm \hat H({\bf k}) + eV({\bf r}, t)}$. Provided the magnetic or
exchange field is constant and uniform, so that the Zeeman term
has no time or space dependence, the basis states ${\rm \{\ket{u_i}\}}$
will depend only on {\bf k}. The reason for this is that all the
spatial and time dependence of the wave functions will only come
from the spatial and time dependence of the vector potential ${\rm{\bf
A} ({\bf r}, t)}$. We will therefore restrict our attention to
constant uniform magnetic fields, while the electric fields may be
space- and time-dependent. As the electromagnetic fields vary on a
spatial scale which is large compared to that of the wave packet,
the local Hamiltonian will have the form ${\rm \hat H[{\bf q}+
\frac{e}{\hbar}{\bf A} ({\bf r}_c, t)] + eV({\bf r}_c, t)}$. The
band eigenstates ${\rm \{\ket{\Psi_{n{\bf k}}}\}}$ take the form
${\rm \ket{\psi_{n{\bf k}}} = e^{i{\bf q}\cdot{\bf r}}\ket{u_{n{\bf
k}}} = e^{i({\bf k} - \frac{e{\bf A}}{\hbar})\cdot{\bf
r}}\ket{u_{n{\bf k}}}}$.  The time dependence of ${\rm \ket{u}}$ comes
both from the Bloch wave functions ${\rm \{\ket{u_i}\}}$, which depend
only on {\bf k}, and from the coefficients, which depend only on
time. Therefore, the Lagrangian in the presence of electromagnetic
fields can be written as:
\begin{equation}\begin{split}
{\cal L} = \hbar\langle u|i\frac{d}{dt}|u\rangle + [\hbar{\bf k}_c
- e{\bf A}({\bf r}_c, t)]\cdot{\bf \dot r}_c - \\ - \epsilon -
\Delta_n - \Delta_d - e V({\bf r}_c, t).
\end{split}\end{equation}
The equations of motion now take the following form: \begin{equation}\begin{split}
\label{eqmo}
\hbar\dot{\bf k}_c = -e({\bf E} + \dot{\bf r}_c\times{\bf B})\el
\hbar\dot{\bf r}_c = \pd{}{{\bf k}_c}\tbkt{u}{\hat H}{u} -
\hbar\dot{\bf k}_c\times{\bf \Omega} + {\bf \Omega}_{t{\bf k}} \el
i\hbar\td{\eta_i}{t} = ({\cal H}_{ij} - \hbar\dot{\bf k}_c\dbkt{u_i}{i\pd{u_j}{{\bf k}_c}})\eta_j,
\end{split}\end{equation} where
${\rm {\bf \Omega} = i\bra{\pd{u}{{\bf k}}}\times\ket{\pd{u}{{\bf
k}}}}$. Note that the position vector equation of motion is very
similar to the one band case \cite{Ganesh}, excepting the presence
of the vector ${\rm {\bf \Omega}_{t{\bf k}}}$, which is non-zero due to
the time dependence of ${\rm \ket{u}}$ through the coefficients. The
equation of motion for ${\rm \ket{u}}$, if a magnetic field is present,
leads to the formula for Larmor precession. The equations may be
solved to any desired order in the external fields and are not
limited to the linear response regime (the fields are weak enough
that they do not induce transitions to remote bands).

\section{The probability amplitudes}
The treatment we have presented so far is an exact analogy with
the single-band dynamics. The equations of motion (\ref{eqmo}) are
complete. Nevertheless, the equations of motion can be made more
explicit in terms of the coefficients ${\rm \eta_i}$, and the non-Abelian
quantities emerging in the process illustrate the gauge structure
of the Hilbert space.

The coefficients $\eta_1$, $\eta_2$ give the composition of the
wave packet in terms of the two bands, and it is natural to think
of them as a vector, $\begin{pmatrix}\eta_1 \\ \eta_2\end{pmatrix}$, which will be
called $\eta$. The connection ${\bf R}$ can be expanded in terms of $\eta$: \beq R^\alpha =
\eta^\dag {\cal R}^\alpha \eta + i\eta^\dag\pd{\eta}{q_\alpha} \,\,\, , {\rm where} \,\,\, {\cal
R}^\alpha_{ij} =\langle u_i|i\frac{\partial u_j}{\partial
q_\alpha}\rangle .\eeq 
We will also introduce the time connection ${\rm {\cal T}_{ij} = \dbkt{u_i}{i\pd{u_j}{t}}}$. The Lagrangian in this picture
takes the form: \bea {\cal L} = i\hbar\eta^\dag\frac{D\eta}{Dt} +
\hbar{\bf q}_c\cdot{\bf \dot r}_c - \eta^\dag \cal H \eta.
\end{eqnarray}
where the covariant derivative with respect to time, defined as ${\rm \frac{D}{Dt} = \frac{d}{d t} - i({\cal T} + \dot{\bf q}_c\cdot{\bf R})}$, has been introduced. Specializing in
electromagnetic fields, we end up with the following Lagrangian:
\begin{equation}\begin{split} {\cal L} = \eta^\dag(i\hbar\frac{D}{Dt})\eta +
[\hbar{\bf k}_c - e{\bf A}({\bf r}_c, t)]\cdot{\bf \dot r}_c - \\ -
\eta^\dag {\cal H} \eta - e V ({\bf r}_c, t).
\end{split}\end{equation}
The equations of motion derived from this electromagnetic
Lagrangian are as follows: \bea \hbar\dot{\bf k}_c = -e({\bf E} +
\dot{\bf r}_c\times{\bf B})\el \hbar\dot{\bf r}_c = \eta^\dag
[\frac{D}{D{\bf k}}, {\cal H}]\eta - \hbar\dot{\bf
k}_c\times\eta^\dag{\bf \cal F}\eta \el
i\hbar\frac{D\eta}{Dt}={\cal H}\eta\eea The covariant derivative
with respect to the wave vector has been introduced, which has the form
${\rm \frac{D}{Dk_\alpha} = \frac{\partial}{\partial k_\alpha} - i{\cal
R}^\alpha}$. The non-Abelian Berry curvature
matrix, ${\rm {\bf \cal F}^\gamma_{ij}}$, is expressed in terms of the
field strength tensor corresponding to the covariant wave vector
derivatives: \beq {\bf \cal F}^\gamma_{ij} =
\frac{1}{2}\epsilon^{\alpha\beta\gamma}{\cal
F}^{\alpha\beta}_{ij}\eeq where \bea {\cal F}^{\alpha\beta}_{ij} =
i[\frac{D}{Dk_\alpha}, \frac{D}{Dk_\beta}]_{ij} = \el = \pd{{\cal
R}^\beta_{ij}}{k_\alpha}-\pd{{\cal R}^\alpha_{ij}}{k_\beta} -
i[{\cal R}^\alpha,{\cal R}^\beta]_{ij} \eea This form, which
includes the non-Abelian correction from the commutator of the
connection matrices, makes evident its gauge covariance with
respect to unitary transformations of $\eta$. The
curvature tensor is antisymmetric under interchange of $\alpha$
and $\beta$, while the indices i and j satisfy ${\rm {\cal
F}^{\alpha\beta}_{ij} = ({\cal F}^{\alpha\beta}_{ji})^*}$.

It is seen from the equations of motion that working in the
coupled-band manifold entails the presence of non-Abelian
quantities such as the modified Berry curvature and gauge
covariant group velocity ${\rm \frac{1}{\hbar}[\frac{D}{D{\bf k}},
{\cal H}]}$, which are corrections to the one band equations of
motion needed to ensure gauge covariance. The matrix ${\cal H}$ is
not necessarily diagonal, as it may include energy gradient
corrections. 

We note that equivalent results can be derived using an argument based on the Ehrenfest theorem, as in the extensive work of Shindou and Imura \cite{Shindou}. 

\section{Constant electric field}
We will examine first the case of a constant uniform electric
field acting on two degenerate bands. We choose a gauge such that
the scalar electric potential need not be included in the
Hamiltonian, and the electric field is represented purely by the
vector potential {\bf A}. With experiment in mind, we take {\bf
E}=(0,0,E), modeling a gate field, and study its effect on
transport in the xy-plane.

\subsection{Electrical spin separation}
We choose as an example the spherical four-band model: 
\begin{equation}\label{HLutt}
\hat H_{Lutt} = \frac{\hbar^2}{2m}[(\gamma_1 +
\frac{5}{2}\gamma_2)k^2 - 2\gamma_2({\bf k}\cdot{\bf \hat J})^2],
\end{equation}
where $\hat{\bf J}$ is the total angular momentum operator, m is the bare electron mass and $\gamma_1$ and $\gamma_2$ are material-specific parameters. The wave functions are eigenstates of the helicity operator ${\bf k}\cdot{\bf \hat J}$ and have the form ${\rm \ket{u_m} = e^{-i\phi J_z}e^{-i\theta J_y}\ket{m}}$ where ${\rm \ket{m}}$ are eigenstates of the angular momentum operator ${\rm J_z}$ while $\theta$ and $\phi$ are the polar and azimuthal angles of the wave vector, respectively. We shall treat the two-fold degenerate heavy and light hole manifolds separately and we shall denote the probability amplitudes in the heavy hole subspace by ${\rm \eta^H}$ and those in the light hole subspace by ${\rm \eta^L}$. 

In these subspaces, the equations of motion for
the probability amplitudes take the form \beq\begin{split} i\hbar\td{\eta^H}{t} = ({\cal H}^H - e E {\cal R}^{zH})\eta^H, \\
i\hbar\td{\eta^L}{t} = ({\cal H}^L - e E {\cal R}^{zL})\eta^L,
\end{split}\eeq 
where the superscripts H and L represent restrictions to the heavy and light hole subspaces, respectively. The reciprocal space connection matrix is given by the following expression:
\beq {\bf \cal R} = \pd{\theta}{{\bf k}}J^y +
\pd{\phi}{{\bf k}}(J^z\cos\theta - J^x\sin\theta). \eeq
In the heavy hole sector ${\rm{\cal R}^z}$=0 and the bands decouple, therefore no spin separation can be achieved electrically in the heavy hole manifold. Henceforth we shall concentrate on the light hole manifold, where the connection matrix
${\rm {\cal R}^z = -\frac{k_\perp}{k^2}\sigma^y}$ has off-diagonal elements only, with
${\rm {\bf k}_\perp = (k_x, k_y)}$ and ${\rm \sigma^y}$ a Pauli spin matrix. We shall suppress the index L in what follows.

The equations of motion for the position and wave vector are: \beq\begin{split}
\hbar\dot{\bf k} = e{\bf E} \el \hbar\dot{\bf r}_c =
\pd{\varepsilon_l}{{\bf k}} - e{\bf E}\times\eta^\dag{\bf \cal
F}\eta,\end{split}\eeq in which ${\rm {\bf k}_0}$ is the initial value of {\bf k},
${\rm \varepsilon_l = \frac{\hbar^2k^2}{2m_l}}$ is the light hole
energy, ${\rm m_l}$ is the light hole effective mass, and the curvature
${\rm {\bf \cal F}=\frac{3}{2}\frac{{\bf k}}{k^3}\sigma^z}$. The wave
vector equation of motion is readily integrated to give ${\rm {\bf k} =
{\bf k}_0 + \frac{e{\bf E}t}{\hbar}}$. Since the Berry curvature vector is
parallel to {\bf k}, there are two limiting cases to consider: the
case ${\bf k}_{\rm 0}//{\bf E}$ is trivial because the curvature
correction vanishes and the bands decouple, so we will focus on
the more interesting case ${\bf k}_{\rm 0} \perp {\bf E}$.

The equations of motion can be solved exactly. $\eta$ is
given by: \beq \eta = \begin{pmatrix}\eta_1^{(0)}\cos\alpha +
\eta_2^{(0)}\sin\alpha \\ \eta_2^{(0)}\cos\alpha -
\eta_1^{(0)}\sin\alpha\end{pmatrix},\eeq with the angle ${\rm \alpha(\tau) =
\arctan(\frac{\tau +
\cos\theta_0}{\sin\theta_0})-(\frac{\pi}{2}-\theta_0)}$, where we
have introduced the dimensionless time ${\rm \tau = \frac{e E t}{\hbar
k_0}}$ and ${\rm \theta_0}$ is the polar angle of ${\bf k}_{\rm 0}$, and where
${\rm \eta_i^{(0)}}$ are the values of $\eta$ at $\tau$=0.

In this system, the contraction ${\rm \eta^\dag {\hat \sigma^i} \eta}$
(with i=1,2,3) is the expectation value of the pseudo-spin.
Its components evolve in time as: \beq\begin{split} \bkt{\hat \sigma^1} =
\bkt{\hat \sigma^1}_{\tau=0}\cos 2\alpha - \bkt{\hat
\sigma^3}_{\tau=0}\sin 2\alpha\el \bkt{\hat \sigma^2} = \bkt{\hat
\sigma^2}_{\tau=0} \el \bkt{\hat \sigma^3} = \bkt{\hat
\sigma^3}_{\tau=0}\cos 2\alpha + \bkt{\hat \sigma^1}_{\tau=0}\sin
2\alpha .\end{split}\eeq The electric field therefore only rotates the 1 and 3
components of the pseudo-spin into combinations of each other,
while the 2 component remains unaffected. To understand the
significance of these results we will examine a concrete example,
taking initially a positive helicity eigenstate so that
${\rm \eta_1^{(0)} = 1, \eta_2^{(0)}=0}$, and fixing the initial wave
vector along the x-axis such that ${\rm {\bf k}_0 = k_0{\bf \hat x}}$,
which means that ${\rm \theta_0 = \frac{\pi}{2}}$. The full time evolution of the pseudo-spin components is:
\beq\begin{split} \bkt{\hat \sigma^1} =
\bkt{\hat \sigma^1}_{\tau=0}\frac{1-\tau^2}{1+\tau^2} - \bkt{\hat
\sigma^3}_{\tau=0}\frac{2\tau}{1+\tau^2} \\ \bkt{\hat \sigma^2} = \bkt{\hat
\sigma^2}_{\tau=0} \\ \bkt{\hat \sigma^3} = \bkt{\hat
\sigma^3}_{\tau=0}\frac{1-\tau^2}{1+\tau^2} + \bkt{\hat \sigma^1}_{\tau=0}\frac{2\tau}{1+\tau^2} .\end{split}\eeq
As $\tau\rightarrow\infty$, $\alpha$ reaches the limiting value of
$\frac{\pi}{2}$ and the components of the pseudo-spin become: \bea
\bkt{\hat \sigma^1} = -\bkt{\hat \sigma^1}_{\tau=0}\el \bkt{\hat
\sigma^2} = \bkt{\hat \sigma^2}_{\tau=0} \el \bkt{\hat \sigma^3} =
-\bkt{\hat \sigma^3}_{\tau=0}. \eea Thus the 1 and 3 components of
the pseudo-spin are reversed while the 2 component is conserved.

The time evolution of the wave vector is described entirely in terms of the time evolution of the angle $\theta$, which is most conveniently expressed as
\begin{equation}\begin{split}
\cos\theta = \frac{k_z}{k} = \frac{\frac{eEt}{\hbar}}{\sqrt{k_0^2 + (\frac{eEt}{\hbar})^2}} = \frac{\tau}{\sqrt{1 + \tau^2}}, \\
\sin\theta = \frac{k_\perp}{k} = \frac{k_0}{\sqrt{k_0^2 + (\frac{eEt}{\hbar})^2}} = \frac{1}{\sqrt{1 + \tau^2}}.
\end{split}\end{equation}
Therefore, initially we have $\cos\theta$=0 and $\sin\theta$=1 while as $\tau\rightarrow\infty$, $\cos\theta\rightarrow$1 and $\sin\theta\rightarrow$0.

The expectation value of a spin component operator ${\rm \hat s^\alpha}$ in the wave packet ${\rm \ket{w}}$ is given by ${\rm \tbkt{w}{\hat s^\alpha}{w} = \eta^\dag s^\alpha \eta}$, where ${\rm s^\alpha_{ij} = \tbkt{u_i}{\hat s^\alpha}{u_j}}$. The time evolution of the spin of one electron can thus be found by knowing the time evolution of its pseudo-spin. Since our goal is to separate spins of opposite orientations, it is sufficient to know the value of the pseudo-spin. The bands being spin-split holes with pseudo-spin up also have spin up and holes with pseudo-spin down have spin down. However, it is instructive to follow the motion of the spin as time progresses, as well as the time evolution of the helicity. The expectation values of ${\rm \hat s^x}$, ${\rm \hat s^y}$ and ${\rm \hat s^z}$ are:
\begin{equation}\begin{split}\label{spinexpval}
\bkt{\hat s^x} = \frac{\hbar}{3}(\frac{1}{2}\sin\theta\cos\phi\bkt{\hat\sigma^3} + \cos\theta\cos\phi\bkt{\hat\sigma^1} - \sin\phi\hat{\sigma^2})\\
\bkt{\hat s^y} = \frac{\hbar}{3}(\frac{1}{2}\sin\theta\sin\phi\bkt{\hat\sigma^3} + \cos\theta\sin\phi\bkt{\hat\sigma^1} + \cos\phi\hat{\sigma^2})\\
\bkt{\hat s^z} = \frac{\hbar}{3}(\frac{1}{2}\cos\theta\bkt{\hat\sigma^3} - \sin\theta\bkt{\hat\sigma^1}).
\end{split}\end{equation}
We assume the carriers have been polarized (by optical means, for example as done in the experiments of Malajovich {\it et al.} \cite{Malajovich, Malajovich2}, although those utilized electrons) so that ${\rm \eta^{(0)}}$ is either $\begin{pmatrix} 1 \\ 0 \end{pmatrix}$ or $\begin{pmatrix} 0 \\ 1 \end{pmatrix}$. Therefore the initial expectation values of ${\rm \{ \hat\sigma^i \}}$ are:
\begin{equation}\begin{split}
\bkt{\hat\sigma^1}_\uparrow = \bkt{\hat\sigma^1}_\downarrow = 0 \\ 
\bkt{\hat\sigma^2}_\uparrow = \bkt{\hat\sigma^2}_\downarrow = 0 \\ 
\bkt{\hat\sigma^3}_\uparrow = - \bkt{\hat\sigma^3}_\downarrow = 1 ,
\end{split}\end{equation}
and the initial spin expectation values are given by
\begin{equation}\begin{split}
\bkt{\hat s^x}_0 = \pm\frac{\hbar}{6}, \bkt{\hat s^y}_0 = \bkt{\hat s^z}_0 = 0.
\end{split}\end{equation}
It can then easily be seen that the y-component of the spin is zero at all times. Substituting for $\theta$ and ${\rm \{ \hat\sigma^i \}}$ in (\ref{spinexpval}) we obtain the time evolution of the other two spin components:
\begin{equation}\begin{split}
\bkt{\hat s^x}_0 = \pm\frac{\hbar}{3}\frac{1 - 5\tau^2}{2(1 + \tau^2)^{3/2}}, 
\bkt{\hat s^z}_0 = \pm\frac{\hbar}{3}\frac{\tau(5 - \tau^2)}{2(1 + \tau^2)^{3/2}}.
\end{split}\end{equation}
As $\tau\rightarrow\infty$, the expectation values of the spin components are:
\begin{equation}\begin{split}
\bkt{\hat s^x} = \bkt{\hat s^y} = 0 \\ 
\bkt{\hat s^z} = \mp\frac{\hbar}{6}.
\end{split}\end{equation}
The spin in this case is not conserved. However, a closer look at (\ref{spinexpval}) reveals that ${\rm \bkt{\hat s^x}}$, ${\rm \bkt{\hat s^y}}$ and ${\rm \bkt{\hat s^z}}$ cannot be obtained from $\bkt{\hat \sigma^1}$, $\bkt{\hat \sigma^2}$ and $\bkt{\hat \sigma^3}$ by a rotation, as the matrix describing the transformation is not unitary. Therefore one should not think of the projection of the spin onto the light-hole subspace as a vector.

Finally, the helicity is given by:
\begin{equation}\begin{split}\label{hel}
\chi = \frac{{\bf k}\cdot\bkt{\hat{\bf s}}}{k} = \frac{k_x\bkt{\hat s^x} + k_z\bkt{\hat s^z} }{k} = \frac{\hbar}{6}\bkt{\hat\sigma^3} = \pm \frac{\hbar}{6}\frac{1 - \tau^2}{1 + \tau^2}.
\end{split}\end{equation}
The helicity is proportional to the expectation value of the third component of the pseudo-spin. It is therefore not conserved for light holes in an electric field. This conclusion was also reached by Jiang {\it et al.} \cite{JiangZF}.

The ${\bf r}_{\rm c}$ equation of motion can be integrated to give the
trajectories of the carriers: \beq\label{rcevol}{\bf r}_c =
\frac{\hbar^2k_0^2}{eEm_l}(\tau {\bf \hat x} +
\frac{\tau^2}{2}{\bf \hat z}) -
\frac{\tau(3+\tau^2)(|\eta_1^{(0)}|^2-|\eta_2^{(0)}|^2)}{2k_0(1+\tau^2)^{3/2}}{\bf\hat
y}.\eeq We have omitted a term proportional to
${\rm \eta_1^{(0)}\eta_2^{(0)}}$ since in our setup either one of them
will be zero . The second term in (\ref{rcevol}) will have opposite signs for the carriers with $\eta$ initially up and those with $\eta$ initially down. Therefore, these carriers will be separated in the y-direction. From the above and Fig. 4.1 it can be seen that the maximum separation in the y-direction occurs at
$\tau=1$ while as $\tau\rightarrow\infty$ this separation tends to ${\rm 1/k_0}$. 

\subsection{Experimental observation}

We discuss an experimental setup in which the effect we have described can be measured. We propose using a three-dimensional semiconductor slab containing a non-degenerate hole gas. The sample must be clean in order for the hole spin relaxation time to be long, specifically of the order of picoseconds. Carriers are excited optically from the conduction band into the valence bands by using a laser beam. Provided the laser beam is sharp, only a narrow range of k-space will be excited around {\bf k}=0. The optically excited holes will have wave vectors lying in a narrow spot about the origin. We assume they have been excited into a state of definite spin. Both light and heavy holes are excited but, as shown in the previous section, the heavy holes do not separate according to spin under the action of the electric field. A source and a drain will be positioned along the x-direction on the two faces of the sample while a gate terminal will be present on top. After the optical excitation, the magnitude of the holes' wave vector can be increased by applying a source-drain field ${\rm E_x}$ in the form of a picosecond pulse, which will accelerate the carriers along the x-axis, its magnitude tuned to ensure ${\rm k_0}$ has the desired value. This source-drain field provides an additional advantage. In the process of optical excitation electrons as well as holes will be excited in the sample and the field which drives the holes one way will drive the electrons the other way, ensuring the effect observed is indeed due to holes. By adjusting the magnitude of the source-drain electric field pulse the initial wave vector ${\rm k_0}$ of the holes incident upon the interface is tunable over several orders of magnitude. We will choose a source-drain electric field in such a way that the wave vector ${\rm {\bf k}_0}$ will have an x-component which overwhelms the y- and z-components. We will also choose the magnitude of ${\rm k_0}$ to be approximately ${\rm 1/b}$, where b is the real space thickness of the laser beam. The reason for this is that in the limit of large $\tau$ the spins are separated by a distance of approximately ${\rm 1/k_0}$, therefore the separation of the spins will be approximately the same as the width of the laser beam. Once excited, the carriers will be subjected to the action of the gate electric field {\bf E} along $\hat{\bf z}$, which will lead to separation of spins as described above. The spin accumulation at the other end of the sample can be measured by Faraday or Kerr rotation. It will be position-dependent along y, that is, as one moves along $\hat{\bf y}$ the spin-z polarization will change sign.

We take the dimensions of the slab to to be 50 nm $\times$ 5 $\mu$m $\times$ 5 $\mu$m and the width of the laser beam is taken as  1 $\mu$m. The optically excited holes will be accelerated until their wave vector reaches the value $k_0 = $10$^6$ m$^{-1}$. For a source-drain field ${\rm E_x}$ of 500 Vm$^{-1}$ and a light hole mass of 0.1${\rm m_0}$, where ${\rm m_0}$ is the bare electron mass, the distance traveled by the light holes along the x-axis will be ${\rm \frac{\hbar^2k_0^2}{2m_leE_x}}$ = 7.2 nm. This will happen after a time of 1.25 ps, which can be achieved in samples in which the holes have longer spin lifetimes. Therefore the source-drain field must be a 1.25 ps pulse of amplitude 500 V m$^{-1}$. 

We will take the gate electric field E=25000 V m$^{-1}$. If one waits for the value of $\tau$ to reach 50, then the magnitude of the spin polarization along the z-direction will be approximately $\hbar/6$ while along the x-direction it will be negligible. The separation between the carriers with spin-z up and spin-z down will be approximately 1 $\mu$m, which is the same as the real-space width of the laser beam and thus observable. The waiting time will be approximately 1.3 ps. Finally, the distances traveled in the x- and z-directions under the action of the gate electric field are 35 nm and 850 nm respectively. 

This phenomenon is similar to effects such as the spin Hall effect since carriers with different helicities are separated in the xy-plane by the electric field normal to the plane.

\begin{figure}[h]
\begin{center}
\epsfxsize=2.5in \epsffile{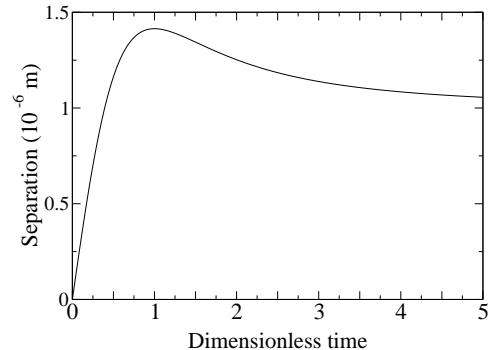} \vspace{0.2cm}
\caption{Separation  l  in the y-direction between light holes
of opposite helicities as a function of $\tau$, the dimensionless
time. }                                          
\end{center}
\end{figure}

\section{Constant magnetic field}
When a constant uniform magnetic field is present, the gradient
correction to the degenerate part of the Hamiltonian takes the form: \beq \Delta_d = -{\bf
M}\cdot{\bf B} .\eeq  {\bf M}, which is identified with the
intrinsic magnetic moment of the wave packet \cite{Chang, Ganesh, Artem}, is given by the expression: \begin{equation}\begin{split}{\bf M} = \pm \frac{e}{2}
\Re\tbkt{u}{{\bf \hat v}\times (i\pd{}{{\bf k}} - {\bf R})}{u} =
\el = \pm \frac{e}{2} \Re(\eta_i^*\tbkt{u_i}{{\bf \hat v}\times (i\pd{}{{\bf
k}} - {\bf R})}{u_j}\eta_j),\end{split}\end{equation} where the sign is negative forelectrons and positive for holes. The operator ${\rm {\bf v} = \frac{1}{\hbar}\pd{\hat H_d}{{\bf k}}}$ is the velocity operator corresponding to the degenerate part of the Hamiltonian. 

Written explicitly in component form and restricting our attention to holes, the magnetic moment is: \beq
M^\alpha =
-\frac{e}{4}\epsilon^{\alpha\beta\gamma}\eta^*_i\tbkt{u_i}{\{\hat
v_\beta , (i\pd{}{k_\gamma} - R^\gamma)\}}{u_j}\eta_j.\eeq
$\epsilon^{\alpha\beta\gamma}$ represents the antisymmetric
tensor. It is straightforward to prove that 
\begin{equation}
\eta_i^* \tbkt{u_i}{\hat{\bf v}}{u_j} \eta_j \equiv \eta^\dag {\bf v} \eta = \frac{1}{\hbar} \pd{\epsilon}{{\bf k}},
\end{equation}
in which ${\rm {\bf v}_{ij} = \tbkt{u_i}{\hat{\bf v}}{u_j}}$. Therefore the second term in {\bf M} is: 
\begin{equation}
{\bf M}_2 = \frac{e}{2} \eta^\dag {\bf v} \eta \times{\bf R} = -\frac{e}{2\hbar} \pd{\epsilon}{{\bf k}}\times{\bf R}.
\end{equation}
The first term is
\begin{equation}
{\bf M}_1 = -\frac{e}{2\hbar} \pd{\epsilon}{{\bf k}}\times{\bf R} - \frac{e}{2}\Re\sum_{i,j}^{in}\sum_l^{out}\eta_i^*{\bf v}_{il}\times{\cal R}_{lj}\eta_j,
\end{equation}
where `out' means the sum runs over all bands outside the degenerate subspace, that is ${\rm l\ne i,j}$. The first contribution exactly cancels ${\bf M}_2$, so the final result is
\begin{equation}
{\bf M} = - \frac{e}{2}\Re\sum_{i,j}^{in}\sum_l^{out}\eta_i^*{\bf v}_{il}\times{\cal R}_{lj}\eta_j.
\end{equation}
Thus the magnetic moment can be expressed purely in terms of matrix elements connecting the degenerate subspace to bands outside the subspace. 

We take as an example once again the light-hole manifold of the
four-band Luttinger model in the spherical approximation in the
presence of a constant uniform magnetic field. The Hamiltonian in
this case is: \beq \hat H = \hat H_{Lutt} -
\frac{ge}{m} {\bf \hat S}\cdot {\bf B}, \eeq where ${\rm \hat H_{Lutt}}$ has been defined in (\ref{HLutt}) and g is the Lande g-factor. The first part of the Hamiltonian is ${\rm \hat H_d}$ while the Zeeman term is ${\rm \hat H_n}$. The Zeeman interaction between the spin and the magnetic field does not contribute to the velocity operator and therefore it does not contribute to the magnetic moment. The light-hole intrinsic magnetic moment in the spherical four-band
model is given by the following expression: \beq {\bf M} =
\frac{3e\hbar\gamma_2{\bf\hat k}}{2m}\bkt{\hat\sigma^3}.\eeq
The magnetic moment is proportional to the expectation value of the third component of the pseudo-spin and therefore to the helicity, as shown in (\ref{hel}).
Depending on the weight of each band in the wave packet the
intrinsic magnetic moment can be positive or negative and if the
bands are equally represented it will be zero.

\end{document}